\documentclass[twocolumn,secnumarabic,amssymb, nobibnotes, aps, prd,superscriptaddress]{revtex4-2}
\usepackage{graphicx}
\usepackage{xcolor}
\usepackage{amsmath}
\usepackage{gensymb}
\usepackage{tabularx}
\usepackage{multirow}
\usepackage[left,modulo]{lineno}
\usepackage{upgreek}

\usepackage{siunitx}

\newcommand{\tableSupplsimuParams}{S-I}
\newcommand{\figSupplConeFieldTime}{S-3}
\newcommand{\figSupplMesureDelta}{S-1}
\newcommand{\figSupplA}{S-5}

\begin{document}
\title{How dirt cones form on glaciers: field observation, laboratory experiments and modeling}

\author{Marceau H\'enot}
\affiliation{Univ Lyon, ENS de Lyon, Univ Claude Bernard, CNRS, Laboratoire de Physique, F-69342 Lyon, France}
\affiliation{SPEC, CEA, CNRS, Université Paris-Saclay, CEA Saclay Bat 772, 91191 Gif-sur-Yvette Cedex, France.}

\author{Vincent J. Langlois}
\affiliation{Laboratoire de Géologie de Lyon, Terre, Planètes, Environnement, Université Claude Bernard Lyon 1 - ENS de Lyon - Université Jean Monnet Saint-\'Etienne - CNRS, France}

\author{Nicolas Plihon}
\affiliation{Univ Lyon, ENS de Lyon, Univ Claude Bernard, CNRS, Laboratoire de Physique, F-69342 Lyon, France}

\author{Nicolas Taberlet}
\email[Corresponding author: ]{nicolas.taberlet@ens-lyon.fr}

\affiliation{Univ Lyon, ENS de Lyon, Univ Claude Bernard, CNRS, Laboratoire de Physique, F-69342 Lyon, France}

\date{\today}
\begin{abstract}
Dirt cones are meter-scale structures encountered at the surface of glaciers, which consist of ice cones covered by a thin layer of ashes, sand or gravel, and which form naturally from an initial patch of debris. In this article, we report field observations of cone formation in the French Alps, laboratory-scale experiments reproducing these structures in a controlled environment, and two-dimensional discrete-element-method–finite-element-method numerical simulations coupling the grains mechanics and thermal effects.
We show that cone formation originates from the insulating properties of the granular layer, which reduces ice melting underneath as compared to bare ice melting. This differential ablation deforms the ice surface and induces a quasi static flow of grains that leads to a conic shape, as the thermal length become small compared to the structure size.
The cone grows until it reaches a steady state in which the insulation provided by the dirt layer exactly compensates for the heat flux coming from the increased external surface of the structure. %This explains the observed long life these structures.
These results allowed us to identify the key physical mechanisms at play and to develop a  model able to quantitatively reproduce the various field observations and experimental findings.

\end{abstract}
\maketitle

%%%%%%%%%%%%%%%%%%%%%%%%%%%%%%%%%%%%%%%%
%%%%%%%%%%%%%%%%%%%%%%%%%%%%%%%%%%%%%%%%
\section{Introduction}
Differential ablation of ice or snow (a disparity in the melting or sublimation rate) is a powerful driving force governing the formation of various natural structures. The mechanism of ablation can be sublimation in the case of blue ice ripples observed in Antarctica~\cite{bintanja_detailed_2001}, elongated snow structures called penitentes found in the Andes mountains~\cite{Mangold2011,Bergeron2006,Claudin2015} or zen stones observed on Lake Baikal which consist of a pebble sitting on a centimetric ice foot caused by an umbrella effect~\cite{ZenStones2021}. Ice melting patterns are observed in various situations: scallops appear at the interface with water under the effect of turbulent flow~\cite{bushuk2019ice,Weady_2022} while suncups form on snow surfaces exposed to solar radiation~\cite{betterton_theory_2001, mitchell_growth_2010}. For the latter, the presence of grains in the snow can play a role in their formation~\cite{rhodes_armstrong_warren_1987}.

%debris covered glacier
The surface of glaciers can be partially or completely covered by a layer of debris (rocks, gravel, ashes, etc) which affects the ablation rate of the ice underneath and has to be taken into account in models attempting to predict the global melt water discharge of glaciers. If thick enough (typically more than 0.5~cm), a debris cover act as an insulation layer and reduces the ice ablation rate. On the contrary, a thin layer enhances the ablation rate compared to a bare ice surface~\cite{ostrem1959}. This later effect has been explained by the patchiness of thin layers~\cite{reid_energy-balance_2010} and by their porosity to air flow~\citep{evatt_glacial_2015} although the lower surface albedo of debris can also play a role, especially in the case of ashes~\cite{rivera2012glacier}. The effect of the presence of a debris layer on the ice ablation rate was well-captured by detailed energy balance models (taking into account the various heat fluxes reaching the surface)~\cite{reid_energy-balance_2010, collier2014representing}, as well as simpler enhanced temperature index models~\cite{carenzo2016enhanced, moeller2016impact} (relying on empirical formulations of the incoming heat fluxes).

%%%%%%%%%%%%%%%%%%%%%%%%%%%%%%%%%%%%%%%%
\begin{figure}[htbp]
  \centering
  \includegraphics[width=\columnwidth]{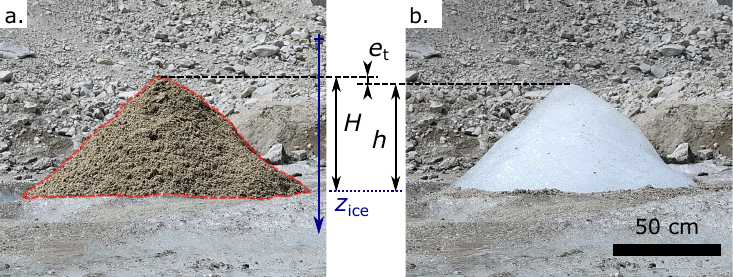}
 \caption{(a) Dirt cones (red dashed contour) of height $H$ observed at the surface of a temperate glacier (Mer de Glace, Alpes). The vertical position of the ice surface is denoted by $z_\mathrm{ice}$. (b) Same structure cleaned of its dirt layer, showing an ice cone of height $h$. $e_\mathrm{t}=H-h$ is the thickness of the dirt layer at the top.}
  \label{fig1}
\end{figure}
%%%%%%%%%%%%%%%%%%%%%%%%%%%%%%%%%%%%%%%%

On temperate glaciers, two type of structures are related to the presence of debris. On the one hand, glacier tables are rocks supported by an ice foot that forms due to a decrease in the melting rate underneath the stone~\cite{Agassiz1840, Bouillette1933, Bouillette1934, OnsetHenot2021, henot2022_TC}. On the other hand, dirt cones  (see fig.~\ref{fig1}) are conical ice structures covered with a thin layer of ashes, grains, or gravel~\cite{Agassiz1840,swithinbank1950origin,wilson1953initiation,krenek_1958,Campbell1975, drewry_1972}. Their height typically ranges from 10~cm to 10~m and they form, depending on their size, over the course of a few days to a few weeks in the ablation zone of glaciers, and they can last for a few months. In 1972, a quantitative field study~\cite{drewry_1972} of natural and artificial cones (triggered by the deposition of patches of gravel), showed the existence of an optimum in grain size (1-10~mm) that maximises the structure formation rate: fine grains are easily washed out by melting water (and possibly rain) while coarse ones do not form an homogeneous protective layer. The author proposed the following qualitative explanation: the thermal protection of the ice by the dirt layer leads, through differential ablation, to the growth of a cone. This causes the debris layer to get thinner as it covers a larger surface area which reduces its protective effect and ultimately causes the decay of the structure. This process is affected by the evolution of the relative slopes of the ice cone and of the debris layer that modulate the slow granular creep flow, and by the fact that the deformation of the layer reduces its shear strength. The complexity of the overall process did not allow for a quantitative comparison with field observations. In 2001, a theoretical study~\cite{betterton_theory_2001} focused on the initial growth of dirt cones on snow by performing a linear stability analysis. The instability results from the adhesion of grains on the snow surface, which causes them to accumulate at the top of the cone, locally reducing snow melting. This model cannot be extended to steady states regimes of ice cones for which the grains do not adhere to the surface and flow along the cone.

In this article, we report a quantitative study of the formation dynamics of dirt cones. We have conducted field observations at Mer de Glace, a temperate glacier in the French Alps, where the formation of dirt cones was monitored over the course of a week. Moreover, we have performed laboratory-scale experiments in simple and well-controlled conditions, and we reproduced the first stage of cone formation. To gain insight on how the deformation of the granular layer is coupled to the evolution of the ice surface, we have developed two-dimensional (2D) numerical simulations taking into account both the quasi-static flow of the granular material forming the dirt layer and the heat transfer across it. Finally, we have developed an analytical model that quantitatively captures the experimental and numerical results and allows a better understanding of the physical processes at play. The article is organized as follows. In Sec.~\ref{sec:matmethods}, we first detail the field observation methods and laboratory experiments, and we provide a description of the 2D numerical model. The results from the field observations, laboratory experiments, and numerical simulations are then presented in Sec.~\ref{sec:results}. A model of cone formation in the laboratory, for which the heat fluxes received by the ice and debris cover can be considered as proportional to the surface temperature of the receiving body is described in Sec.~\ref{sec:model} and accurately predicts the evolution of the dirt-cone observed in the laboratory. This model is then refined to take into account solar radiation and albedo of the ice and of the debris, which successfully reproduces the field observations. Conclusion and perspectives are then presented in Sec.~\ref{sec:ccl}.

\section{Materials and methods}\label{sec:matmethods}

\begin{figure}[htbp]
  \centering
  \includegraphics[width=\columnwidth]{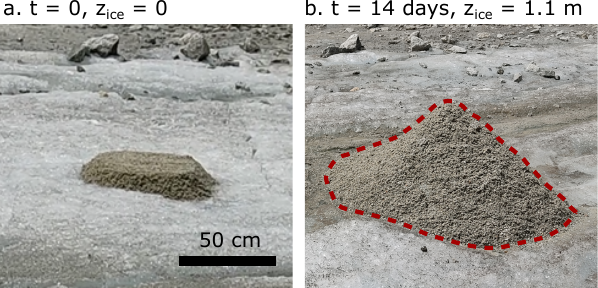}
 \caption{Formation of a dirt cone. (a) Initial gravel pile (artificially made). (b) Cone formed after 14 days (red contour).}
  \label{fig2}
\end{figure}

\textbf{Field observations} were performed on the Mer de Glace glacier in the French Alps at an altitude of 2000~m located at 45\degree54'48.8''~N, 06\degree56'10.9''~E. To follow the initial formation of a cone, we built three initial piles using small gravel (millimetric grain size) found on the side of the glacier. Grains were compacted into circular shapes of radius $R_0$ and uniform thickness $e_\mathrm{t0}$ (see Fig.~\ref{fig2}a, schematics in Fig.~\ref{fig8} and values of the parameters in Table~\ref{tableConesIni}). The evolution of the piles was followed using time-lapse images produced by an autonomous solar-powered camera (Enlaps Tikee), positioned on three 1.5~m-long wood rods set into the ice. Pictures (4608~px $\times$ 3456~px) were taken every 1~h between 5~a.m. and 10~p.m. between June 7 and June 19 2019 (see Figs. 1 and 2), until the camera fell on the ice due to the melting around the supporting rods. The residual motion of the device was corrected by tracking two fixed points on the background of each image. The positions of the top of the cone and of the bottom of its left and righ sides were then manually pointed out on each image (see video C in suppl. mat.~\cite{Suppl}). The air temperature $T_\mathrm{air}$ (3~m above ground), solar radiative flux $\Phi$, and wind speed $u_\mathrm{air}$ were measured at the Requin automatic weather station (AWS)~\citep{nadeau2009estimation} located 600~m higher and 3~km away from the measurement site (see Fig.~\figSupplConeFieldTime~b-d  and ref.~\cite{henot2022_TC} for a discussion on the validity of the assumptions made to compute the local temperature, solar radiative flux, and wind speed). 

%%%%%%%%%%%%%%%%%%%%%%%%%%%%%%%%%%%%%%%%
%%%%%%%%%%%%%%%%%%%%%%%%%%%%%%%%%%%%%%%%
%\VL{Juste redécoupé un peu pour raccourcir les phrases.}

\textbf{Small-scale experiments} were performed in a laboratory-controlled environment protected from parasite air flow and held at constant temperature $T_\mathrm{room} = 25.5$~\degree C. The granular media consisted in a plastic blast media purchased from Guyson, made of 66-70~\% urea amino polymer and 33-30~\% cellulose (density 1.5) with irregular shapes and size lying between 0.84 and 1.20~mm (16/20 mesh size). In order to prevent cohesion and to minimize the thermal conductivity by avoiding water absorption by the medium, these grains were made hydrophobic using a two steps coating with Rust-Oleum NeverWet multisurface spray. Clear ice blocks (cylinders of diameter 30 cm  and height 20 cm) were obtained through unidirectional freezing inside a container thermally isolated on its sides and bottom, and placed inside a -35~\degree C freezer for 3 days. In order to ensure a homogeneous temperature $T_\mathrm{ice}=0$~\degree C, the ice block was then left at ambient temperature before the beginning of the experiments. 

\begin{figure}[htbp]
  \centering
  \includegraphics[width=\columnwidth]{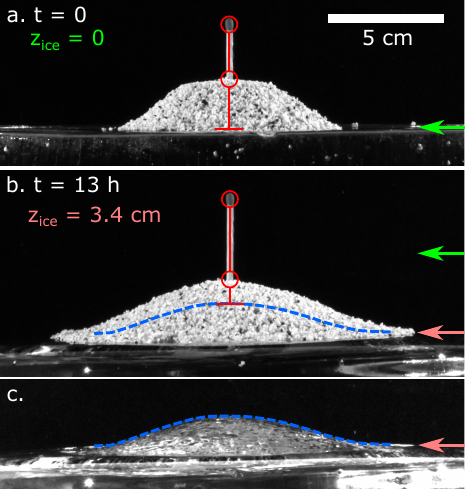}
 \caption{Formation of an artificial dirt cone in a laboratory-controlled environment. A vertical stick (red) allows to measure the tickness of the granular layer at the summit of the cone, and to therefore follow the vertical position of the ice surface at the center of the structure (horizontal red line). (a) Initial pile of plastic granular media on a flat ice surface. (b) Cone formed after 13~h. (c) Same cone with its granular cover removed, revealing the surface of the ice (blue line).}
  \label{fig3}
\end{figure}

For each experiment, a flat pile of grains was deposited at the surface of the ice block (see Fig.~\ref{fig3}a).
The altitude of the ice far from the cone, $z_\mathrm{ice}(t)$, and the height of its summit, $H(t)$, were monitored. 
At the center of the pile, a wooden stick was mounted on a small plastic foot lying on the ice surface, allowing us to measure the thickness of the granular layer $e_\mathrm{t}(t)$. It was checked at the end of the experiment that the plastic foot did not penetrate into the ice surface more than 0.5~mm. The pile was illuminated from the sides using two LEDs, and its evolution was followed by taking a picture every 6~min using a D5600 Nikon with a 200~mm lens placed 3.5~m away from the system. Both the 3D field configuration and the 2D numerical configuration (see below) were reproduced: 3D structures were obtained from an initial circular pile of thickness $e_0$, radius at the bottom, $R_0$, and angle of repose, $\theta_0=36.0 \pm 2.5$\degree. Pseudo-2D structures were obtained from initial rectangular piles of half-width $R_0$ and length $ 4 R_0$. The duration of the experiments was constrained by the melting of the edges of the ice block that limited the maximum lateral extension of the cones.

%%%%%%%%%%%%%%%%%%%%%%%%%%%%%%%%%%%%%%%%
%%%%%%%%%%%%%%%%%%%%%%%%%%%%%%%%%%%%%%%%

%%\VL{Enlevé qq détails... est-ce que les paramètres des simus ne devraient pas aller en annexe plutôt que supp.mat ? }

\begin{figure}[htbp]
  \centering
  \includegraphics[width=\columnwidth]{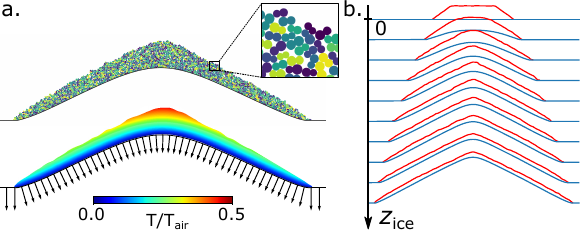}
 \caption{(a) Principle of the 2D numerical simulations. The granular quasi-static creep flow is determined using a soft-disk discrete element method (DEM) (top). The temperature distribution in the media is computed using a finite element method (bottom). The displacement of the ice surface $v_\mathrm{c}$ is locally proportional to the heat flux ($j_\mathrm{c}$, black arrows). (b) Results of a 2D numerical simulation for an initial piles of thickness $e_\mathrm{t0}$ equal to the thermal length $\delta=\lambda/h_\mathrm{eff}$. Only the interfaces (ice/grains (in red) and grains/air (in blue)) are shown.}
  \label{fig4}
\end{figure}

\textbf{2D numerical simulations} were performed by combining the discrete element method (DEM) to model the granular mechanics with a finite element method to compute the thermal fluxes. 
The granular media is modelled as an assembly of 2D deformable disks of average radius $\langle r\rangle = 0.25$~mm (with 20~\% polydispersity), assembled in dimers by adding a constant attractive force within a pair of grains (see Fig.~\ref{fig4}a, top). The use of dimers is a way to mimic grains of aspect ratio larger than 1, which helps to reach a higher angle of repose~\cite{zhou2014angle}.
In addition, disks experience gravity and contact forces (normal inelastic repulsion and frictional tangential force). From the sum of all forces and torques acting on each disk, its translational and rotational motion is computed at each timestep by classical granular DEM techniques~\cite{Poeschel2005}. The values of all numerical parameters are summarized in Table~\tableSupplsimuParams and Table~S1. The initial state consists in a trapezoidal pile of thickness $e_\mathrm{t0}$, half width $R_0$ and angle $\theta_0$ close to the angle of repose of the grains (see Fig.~\ref{fig8}) with $e_\mathrm{t0}/\langle r\rangle \in[40, 120]$ and $R_0/\langle r\rangle \in[80, 240]$. This pile lies on an initially horizontal layer of fixed grains (of size $0.6\langle r\rangle$), representing the first layer of grains glued to the ice surface. At each time step $dt$, each of these fixed grains (\textit{i.e.}, the local position of the ice surface) is moved downwards by a distance $v_\mathrm{c}\times dt$, which varies along the pile as it results from the heat flux within the uneven granular layer. Outside the granular pile, the bare ice surface moves at the ablation velocity $v_\mathrm{ice}$, whose value is small enough to lead to a quasi-static granular flow. Below the pile, ice melting is controlled by the heat flux through the granular layer, treated as an effective medium of thermal conductivity $\lambda$ exchanging heat with air at temperature $T_\mathrm{air}$ with an effective heat exchange coefficient $h_\mathrm{eff}$ and in contact with melting ice at temperature $T_\mathrm{ice}$. The heat flux $j_\mathrm{c}$ delivered to the ice is computed every $5\times 10^4$ time-steps by solving the heat equation with the finite element solver FreeFem++~\cite{MR3043640} (see Fig.~\ref{fig4}a, bottom). The ice velocity below the pile is then $v_\mathrm{c} = v_\mathrm{ice}\times j_\mathrm{c} /(h_\mathrm{eff}(T_\mathrm{air}-T_\mathrm{ice}))$.
%%%%%%%%%%%%%%%%%%%%%%%%%%%%%%%%%%%%%%%%
%%%%%%%%%%%%%%%%%%%%%%%%%%%%%%%%%%%%%%%%

\section{Results} \label{sec:results}

%%%%%%%%%%%%%%%%%%%%%%%%%%%%%%%%%%%%%%%%
\begin{figure}[htbp]
  \centering
  \includegraphics[width=\columnwidth]{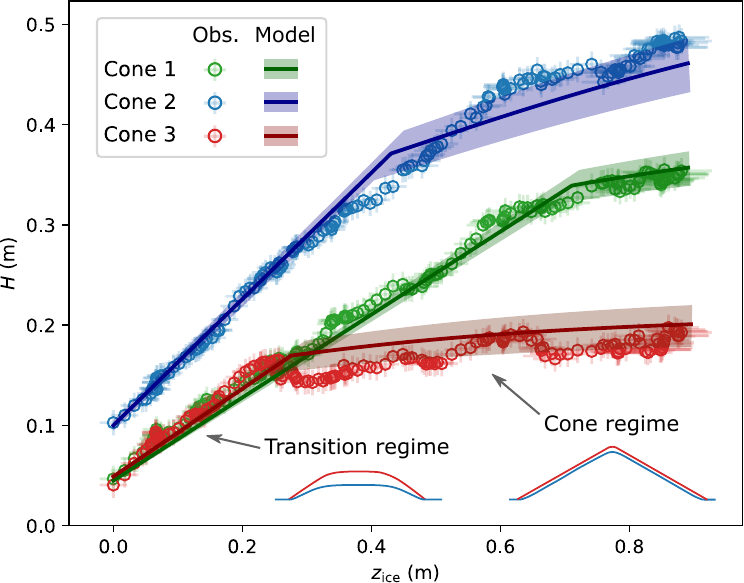}
 \caption{Cone formation dynamics on the Mer de Glace for the 3 gravel piles described in Table~\ref{tableConesIni}. The total cone height $H$ is shown as a function of the total ice ablated thickness $z_\mathrm{ice}$ (markers). The model is plotted in solid line for each initial state ($e_\mathrm{t0}$, $V_0$, the shaded area corresponds to the uncertainty on these parameters) with adjustable parameter $f = 1.4$.}
  \label{fig5}
\end{figure}
%%%%%%%%%%%%%%%%%%%%%%%%%%%%%%%%%%%%%%%%
%\VL{Surtout réordonné les phrases}

In the field and 3D laboratory experiments, initially flat piles of granular media turned into conical structures (see Fig.~\ref{fig2}) after the bare ice was ablated by a thickness $z_\mathrm{ice}\gg e_\mathrm{t0}$ (far away from the structure). Removing the grain cover showed that these \textit{dirt cones} consist in ice cones covered by a thin layer of grains (see Fig.~\ref{fig1}b). The 2D laboratory experiments and numerical simulations displayed a similar behavior: the cross section of the ice surface went from flat to triangular (see Fig.~\ref{fig4}b). Both in 2D and 3D, the evolution of the structure can be divided into two stages, clearly visible in the numerical simulation of Fig.~\ref{fig4}b (see also video B in suppl. mat.): first, a transient regime which lasts until the flat region at the center of the pile disappears and the shape becomes conical (or triangular), then a stable regime in which the cone keeps growing while keeping a constant slope. It is visible in the film that the grain flow is not limited to avalanches at the surface but also takes place in the bulk of the cover layer. The quantities used in the following to quantitatively characterize the structure evolution are shown in Fig.~\ref{fig1}: $h$ denotes the height of the ice dome, $e_\mathrm{t}$ is the thickness of the grain layer at the top and $H=h+e_\mathrm{t}$ the total height. In the cone regime, the slope of the ice cone and that of the dirt cone differ by only a few degrees~\cite{drewry_1972}: in the following this difference will be neglected and a unique value $\theta$ is used. This value strongly depends on the nature of the granular medium: the measured values are displayed in Table~\ref{tableConesAngles} and range from 19\degree~for dry, low friction grains in the simulation, to 49\degree~for wet gravel in the field. Let us note that these angles are systematically lower than the repose angle of the grains by 5 to 10\degree. % and systemically smaller than the repose angle.% In the lab experiments: $\theta=25\pm 5$° (only the beginning of the cone regime is accessible which make the cone angle difficult to measure precisely) while the angle of repose is $\theta_0=36.0 \pm 2.5$°. In the simulation $\theta = 26 \pm 1$° and $\theta_0 = 35 \pm 1$°. On the field the cone angle is $\theta=49\pm 4$° and the initial angle is $\theta_0 \approx 55$° (the grains being wet and cohesive, the repose angle is not well defined).

%%%%%%%%%%%%%%%%%%%%%%%%%%%%%%%%%%%%%%%%
\begin{figure}[htbp]
  \centering
  \includegraphics[width=\columnwidth]{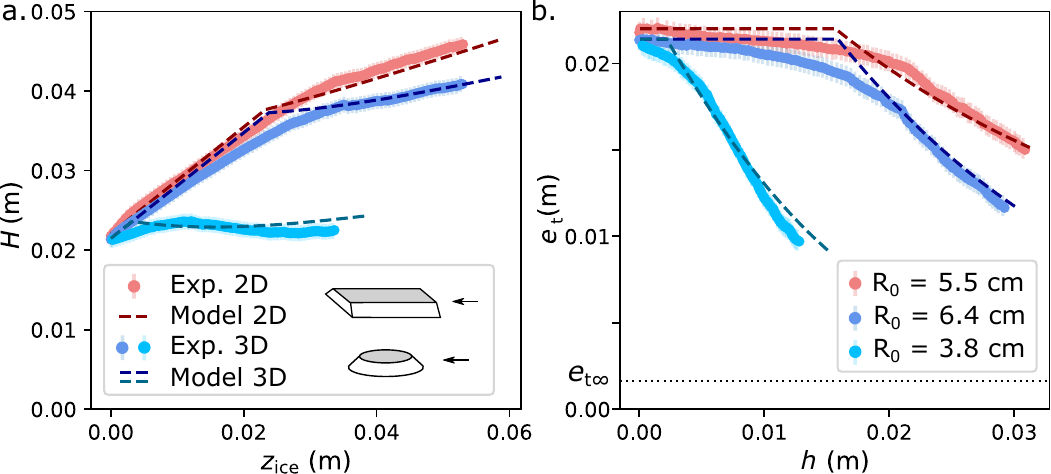}
 \caption{(a,b) Cone formation in the laboratory-controlled environment (markers) in a 2D (red) and 3D (blue) configurations. The total cone height $H$ is shown as a function of the ablated ice thickness $z_\mathrm{ice}$ (a) and the granular thickness as the top of the cone $e_\mathrm{t}$ as a function of its height $h$ (b). The model (see section IV) is plotted in dashed line for the corresponding geometry (2D/3D) and for each initial state ($e_\mathrm{t0}$, $R_0$) with adjustable parameter $\theta = 23.5$\degree.}
  \label{fig6}
\end{figure}
%%%%%%%%%%%%%%%%%%%%%%%%%%%%%%%%%%%%%%%%

\begin{figure}[htbp]
  \centering
  \includegraphics[width=\columnwidth]{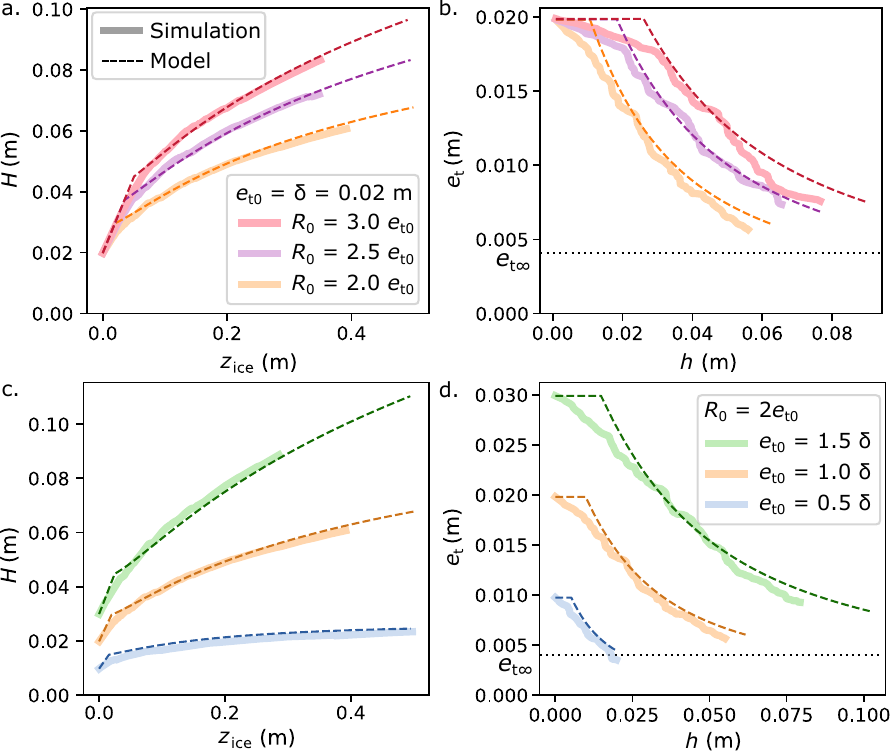}
 \caption{Results of 2D numerical simulations (solid lines), for $\mu=0.6$. The quantities plotted are the same than in Fig.~\ref{fig6}. The model is plotted in dashed line for each initial state ($e_\mathrm{t0}$, $R_0$) with adjustable parameter $\theta = 27$\degree.}
  \label{fig7}
\end{figure}

\begin{figure}[htbp]
  \centering
  \includegraphics[width=\columnwidth]{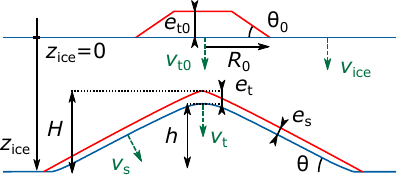}
 \caption{Schematics of an initial pile of grains (top) and of the cone forming when the ice surface melts (bottom). The blue curve corresponds to the ice surface and the red curve to top of the granular cover.}
  \label{fig8}
\end{figure}

In our field observations, only the evolution of the total height $H$ with time (and therefore with $z_\mathrm{ice}=v_\mathrm{ice}\, t$) was accessible and is plotted in Fig.~\ref{fig5}. In numerical simulations and laboratory experiments, all quantities $h$, $e_\mathrm{t}$ and $H$ could be monitored: Figs.~\ref{fig6} and \ref{fig7} show $H(z_\mathrm{ice})$ and $e_\mathrm{t}(h)$ for different initial pile shapes. All these data qualitatively display the same behavior: first, the growth rate of the cone height $\mathrm{d}H/\mathrm{d}z_\mathrm{ice}$ is maximum during the transient regime. For the largest initial radius, this rate keeps a constant value for a little while, meaning that the height first evolves linearly. This is particularly visible on the field data. In this first stage, in the laboratory and in the simulations, the thickness at the summit, $e_\mathrm{t}$, diminishes only slightly ($\lessapprox 10 \%$). Once the cone regime is reached, the growth rate of the cone strongly decays, while the granular cover on top quickly thins down. At long times (\textit{i.e.}, for $v_\mathrm{ice}\,t \gg R_0$), we observe in the field experiments (red markers in Fig.~\ref{fig5}) and in the simulations (blue line in Fig.~\ref{fig7}b) that the growth rate tends to zero. This corresponds to a cone that dynamically keeps the same shape and size while its internal ice is melting at the same rate as the bare ice surface. In the laboratory experiments, this last regime is not accessible due to the limited size of the ice blocks. %\MH{En fait ce point était déjà mentionné à la fin de la p. 6.}\NP{c'est bien de le mentionner ici aussi vu les questions des referees.}

%%%%%%%%%%%%%%%%%%%%%%%%%%%%%%%%%%%%%%%%
%%%%%%%%%%%%%%%%%%%%%%%%%%%%%%%%%%%%%%%%

%%%%%%%%%%%%%%%%%%%%%%%%%%%%%%%%%%%%%%%%
\begin{table}%  table1,  one  column
\caption{Initial characteristics of the cones studied in the field. $V_0$ is computed using Eq.~\ref{eq_suppl_vcone} with $\theta_0=55\pm10\degree$.}
\label{tableConesIni}
\begin{tabularx}{0.38\textwidth}{c 
>{\centering\arraybackslash}X
>{\centering\arraybackslash}X
>{\centering\arraybackslash}X}
\hline
  %\hline
  Cone & $R_0$ & $e_\mathrm{t0}$ & $V_0$ \\
  index & (cm) & (cm) & (L) \\
  \hline
  \hline
   1 & $24 \pm 2$ & $4.5 \pm 0.2$ & $7 \pm 1$ \\
   2 & $26 \pm 2$ & $10.0 \pm 0.5$ & $19 \pm 4$ \\
   3 & $12 \pm 1$ & $4.9 \pm 0.2$ & $1.5 \pm 0.3$ \\
      \hline
\end{tabularx}
\end{table}
%%%%%%%%%%%%%%%%%%%%%%%%%%%%%%%%%%%%%%%%
%%%%%%%%%%%%%%%%%%%%%%%%%%%%%%%%%%%%%%%%
\begin{table}%  table1,  one  column
\caption{Cone angle $\theta$ measured from pictures or profiles and angle $\theta$ and decompaction factor $f$ used in the model. The $\star$ symbol denotes the parameter kept adjustable in the model.}
\label{tableConesAngles}
\begin{tabularx}{0.38\textwidth}{c 
>{\centering\arraybackslash}X
>{\centering\arraybackslash}X
>{\centering\arraybackslash}X}
\hline
  %\hline
  Context & $\theta$ meas. & $\theta$ model & $f$ model\\
  \hline
  \hline
   Lab  & 25\degree $\pm$ 5\degree& 23.5\degree ($\star$) & 1 \\
   \hline
    Simulations  &  &  \\
   $\mu=0.6$  & 26\degree $\pm$ 2\degree & 27\degree ($\star$) & 1 \\
   $\mu=0.3$ & 19\degree $\pm$ 2\degree & 19\degree ($\star$) & 1 \\
   \hline
   Field & 49\degree $\pm$ 4\degree & 49\degree & 1.4 ($\star$) \\
      \hline
\end{tabularx}
\end{table}
%%%%%%%%%%%%%%%%%%%%%%%%%%%%%%%%%%%%%%%%

\section{modeling and discussion.} \label{sec:model}

%%%%%%%%%%%%%%%%%%%%%%%%%%%%%%%%%%%%%%%%
%%%%%%%%%%%%%%%%%%%%%%%%%%%%%%%%%%%%%%%%

\subsection{Cone formation in the laboratory and in simulations}
\label{subsection_cone_lab}

In the following, we develop a simple model of the formation of a cone from a initial flat pile of grains. We first concentrate on the simpler case in which all the heat fluxes coming from the environment can be considered as proportional to the surface temperature of the receiving body. As discussed below, this applies well to our laboratory experiments and was implemented in the numerical simulations presented in this work. On a natural glacier, however, the process is also affected by direct solar radiation and this case will be treated in the next subsection.

\textbf{Ice melting.} The ice surface away from the granular pile gets lower due to melting under the effect of a positive net incoming heat flux $Q_\mathrm{env.\rightarrow ice}$ coming from the environment. Its vertical position $z_\mathrm{ice}(t)$ can be expressed, assuming that the melting process is instantaneous, as $z_\mathrm{ice}(t) - z_\mathrm{ice}(0) = \mathcal{L}_\mathrm{fus}\int_0^t Q_\mathrm{env.\rightarrow ice}(t) \mathrm{d}t$, where $\mathcal{L}_\mathrm{fus}=303$~MJ$\cdot$m$^{-3}$ is the volumetric enthalpy of fusion for ice. In a laboratory-controlled environment, the incoming heat flux has two main origins~\cite{OnsetHenot2021}: the net infrared radiation coming from the enclosure walls (at temperature $T_\mathrm{room}$) and the natural convection of air (also at temperature $T_\mathrm{room}$). Since $T_\mathrm{room}-T_\mathrm{ice} \ll T_\mathrm{ice}$, its expression can be linearized as
\begin{align}
    Q_\mathrm{env.\rightarrow ice}=h_\mathrm{eff}(T_\mathrm{room}-T_\mathrm{ice})
\end{align}
where $h_\mathrm{eff} = 8 \pm 2$~W$\cdot$K$^{-1}\cdot$m$^{-2}$ is an effective heat exchange coefficient that was measured by monitoring the melting of an ice block (see Supp. Mat.).

\textbf{Early stage of the transition regime.} At the very beginning of the process, the structure consists in a flat pile of grains that acts as an insulating cover: the dirt surface being warmer than the ice, it receives less heat from the environment, and therefore $Q_\mathrm{dirt\rightarrow ice}<Q_\mathrm{env.\rightarrow ice}$. 

As long as $R_0 \gg e_\mathrm{t0}$ the process, at the center of the pile, can be considered as one-dimensional (1D), which leads to a simple analytical formulation. Assuming that the effective heat exchange coefficient $h_\mathrm{eff}$ is the same for the dirt and the ice surfaces, the ratio of melting velocity between covered and bare ice is, at $t=0$:
\begin{align}
\frac{v_\mathrm{t0}}{v_\mathrm{ice}} = \frac{Q_\mathrm{dirt\rightarrow ice}}{Q_\mathrm{env.\rightarrow ice}} =\frac{1}{1+e_\mathrm{t0}/\delta}
\label{eq_vt_vice_0}
%\bigg\rvert_{\Phi=0}
\end{align}
where $\delta = \lambda/h_\mathrm{eff}$ is a thermal characteristic length and $\lambda$ is the effective thermal conductivity of the dirt layer ($\mathrm{Bi}=e_\mathrm{t0}/\delta$ is the Biot number). This differential ablation of ice leads to the growth of an ice foot under the dirt pile (in the referential of the bare ice surface) at a rate:
\begin{align}
\frac{\mathrm{d}h}{\mathrm{d}z_\mathrm{ice}} = 1 - \frac{v_\mathrm{t}}{v_\mathrm{ice}}
\label{eq_h_vt_vice_0}
\end{align}

\noindent where $v_\mathrm{t}$ is the vertical melting velocity below the center of the structure. In the early stages, the top of the pile remains flat: there is no driving force leading the grains to move laterally and we can assume that $e_\mathrm{t} \approx e_\mathrm{t0}$. Combining eqs.~(\ref{eq_vt_vice_0}) and (\ref{eq_h_vt_vice_0}) leads to 
\begin{align}
H=h+e_\mathrm{t}=\frac{\mathrm{Bi}}{1+\mathrm{Bi}}\times z_\mathrm{ice}
\label{eq_H_lin}
\end{align}
Using the measured value of the thermal length for the plastic grains $\delta = 10.8\pm 0.3$~mm (measured in an independent experiment, see Fig.~\figSupplMesureDelta), this prediction is plotted (with no adjustable parameter) in Fig.~\ref{fig6}a as dashed straight lines. For the wider piles ($R_0 = 5.5$ and 6.4~cm), for which the 1D approximation is most valid, the prediction fits the observations for $0<z_\mathrm{ice}<2$~cm. For the smallest pile ($R_0 = 3.8$~cm, light blue), the model overestimates the initial growth rate and the linear growth regime is not observed. But even large piles eventually reach a point where the assumptions made above are no longer valid: the grains at the center are affected by the lateral flow induced by the deformation of the sides. As a consequence $e_\mathrm{t}$ decreases and the thermal problem is not 1D anymore. This induces a complex dynamics that ultimately leads to the formation of a conic structure and which is only described qualitatively here. Differential ablation deforms the ice surface at the periphery of the pile, which induces a quasi-static flow in the grain cover whose free surface adopts a slope $\theta_\mathrm{grains}$. The flow modifies the cover thickness and couples back with the deformation of the ice surface, which takes a slope $\theta_\mathrm{ice} < \theta_\mathrm{grains}$ (see Fig.~\ref{fig4}b.). As the cover is thinner on the outside, the ice melts faster on the outside, causing the two angles to converge to the same value $\theta$. The conic shape is obtained when the deformation reaches the center. The fact that the final structure is a cone rather than a smooth dome results from the fact that as the structure grows, its typical dimensions (radius and height) both exceed the thermal length $\delta$ ($\approx 11$~mm) which controls the scale over which the ice profile can vary. In the lab, due to the experimental constraints on the size of the ice block, $\delta$ was made as small as possible by using rather insulating grains, yet the ratio $H/\delta$ is still significantly smaller than in the simulations or on the field. This explains (alongside with the difference in $\theta$) the different shape of the ice cones obtained in the laboratory (fig.~\ref{fig3}c: smooth with low angle), in simulations (fig.~\ref{fig4}b: conic with low angle) and in the field (fig.~\ref{fig1}: conic with large angle).

%\VL{J'ai déplacé tout ce qui concerne l'état initial du cône ($h_1$, $z_1$ etc) au début plutôt qu'après l'équadiff, sinon ça fait des aller-retours avec le régime de transition}

\textbf{Cone regime.} 
In the following we focus on a structure that has already reached the cone regime, (\textit{i.e.}, $e_\mathrm{t},e_\mathrm{s}\ll H$). Let us assume that the approximations $e_\mathrm{t}\approx e_\mathrm{t0}$ and $v_\mathrm{t}\approx v_\mathrm{t0}$ (given by Eq.~\ref{eq_vt_vice_0}) are valid throughout the whole transition phrase: the cone regime is then reached for a total ablation $z_\mathrm{ice}=z_1=h_1/(1-v_\mathrm{t0}/v_\mathrm{ice})$, where $h_1$ is the initial height of the cone at the end of the transient regime. Considering that the total volume $\mathcal{V}$ of the granular cover is conserved, $h_1$ can be expressed (in 2D or 3D) as a function of initial parameters $\theta_0$, $R_0$ and $e_\mathrm{t0}$ (see Eq.~\ref{eq_Ap_h1_2D} and~\ref{eq_Ap_h1_3D} in the appendix).

In order to develop a simple analytical model of the later stage, the following assumptions are made: 

(i) The thermal problem can be considered 1D for the dirt on the sides of the cone: $v_\mathrm{s}/v_\mathrm{ice} = 1/(1+e_\mathrm{s}/\delta)$. This is valid far away from the transition regime, when $e_\mathrm{s}\ll h/\tan \theta$. 

(ii) The angle of the cone, as well as the shape of the top of the ice cone are stationary. This is supported by the numerical simulations (see Fig.~\ref{fig4}b). From this, we can relate the melting velocity at the summit to that on the side of the cone: $v_\mathrm{t}=v_\mathrm{s}/\cos \theta$. 

(iii) The ratio $A = e_\mathrm{s}/e_\mathrm{t}$ is a constant. Although not obvious (since this ratio results from the granular flow and the melting on top of the cone), this assumption is supported by the numerical simulations where $A$ is observed to be constant and independent of the cone angle (for $\theta = 19$ \degree and 26\degree): $A = 0.6 \pm 0.1$. In laboratory experiments and in the field $A$ can only be measured at the end of the evolution, but this final value is equal to the numerical value and independent of the initial shape of the pile (see {Fig.~\figSupplA}). 

From these assumptions the growth rate of the cone height can be expressed as a function of $e_\mathrm{t}(h)$:
\begin{align}
    \frac{\mathrm{d}h}{\mathrm{d}z_\mathrm{ice}} = 1-\frac{1}{\cos \theta}\times \frac{1}{1+A e_\mathrm{t}(h)/\delta}
    \label{eq_dHdzice}
\end{align}

The state at the end of the transient regime is defined by $z_\mathrm{ice}=z_1$, $h(z_1)=h_1$ and $e_\mathrm{t}(h_1)=e_\mathrm{t0}$. Through volume conservation, we express the quantity $e_\mathrm{t}$ both in the 2D and 3D cases (see appendix). Finally we solve Eq.~\ref{eq_dHdzice} numerically with $\theta_0$, $R_0$, $e_\mathrm{t0}$, $A$ and $\delta$ as input parameters. We keep $\theta$ as the only adjustable parameter, due to the high sensibility of the model to this quantity. Furthermore, the value of $\theta$ results from a complex feedback between the evolution of the ice surface and the creep flow in the granular cover, which prevents us from predicting a simple \textit{a priori} estimate.

The best fitting resulting evolution is shown for laboratory experiments in Fig.~\ref{fig6} ($\theta=23.5$\degree) and for numerical simulations in Fig.~\ref{fig7} ($\theta=27$\degree ~and $\mu = 0.6$) (see suppl. mat. for results concerning $\mu=0.3$). The values of the adjustable parameter $\theta$ (see Table~\ref{tableConesAngles}) are within the range of the values that were measured independently.

The good agreement in the $e_\mathrm{t}(h)$ plots of Figs.~\ref{fig6}b (in 2D and 3D) and \ref{fig7}b-d (in 2D) supports in retrospect the assumption of volume conservation in the cone regime. This also shows that the difference in the cone formation dynamics between the 2D and 3D cases is mainly related to the volume conservation. This justifies the relevance of the 2D simulations in testing the other assumptions (i-iii)  of the model, related to mechanical and thermal processes. Our description of the transition regime (constant $e_\mathrm{t}$) is too simplistic: since $e_\mathrm{t}$ shows a perceptible decrease, the model overestimates the growth rate $\mathrm{d}h/\mathrm{d}z_\mathrm{ice}$ but underestimates $z_1$. Remarkably, these errors compensate, which leads to a good agreement between the prediction and the observed evolution $H(z_\mathrm{ice})$ in the cone regime.

\textbf{Steady state.} At long times, Eq.~\ref{eq_dHdzice} predicts the existence of a steady state in which $e_\mathrm{t,\infty} = \delta(1/\cos\theta-1)/A$. This value is independent of the initial conditions (which is not the case of the steady state height and radius of the cone) and is represented in Fig.~\ref{fig6}b and \ref{fig7}b,d using dotted horizontal lines. One can see that only the numerical simulation with the smallest initial thickness approaches its final state. In the laboratory experiments, it was not possible to reach the steady state since,  given the low cone angle, the finite size of the ice block was quickly limiting the maximum value of $z_\mathrm{ice}$.

\subsection{Natural cone formation on a glacier}

The formation of natural dirt cones occurring at the surface of a glacier is slightly more complex than the process taking place in the well-controlled laboratory conditions, mainly because the heat flux coming from the environment cannot simply be described using an effective heat exchange coefficient. Indeed, in the previous model the heat flux received by the ice and dirt surfaces is governed by their temperatures. This is not the case for the solar heat flux which plays a crucial role in the field. In the following we show that the previous model can be adapted to these conditions with only minor modifications.

\textbf{Ice melting.} In the field, the main heat source is direct solar irradiation. For our field data, it represented 60~\% of the total incoming flux, the rest coming from the turbulent fluxes (convection and sublimation or condensation due to the wind), whereas the net infrared radiation was almost null (but negative)~\cite{henot2022_TC}. While it is possible to model in details these physical processes, another approach classically used in glaciology is to rely on an empirical relation known as an enhanced temperature index model~\cite{pellicciotti2005enhanced, carenzo2016enhanced, moeller2016impact}. Let us assume that all contributions other than net solar radiation can be described by an empirical term proportional to the difference between air and surface temperature:
\begin{align}
    Q_\mathrm{env.\rightarrow ice}= (1-\alpha_\mathrm{ice})\Phi(t) + h_\mathrm{eff}(\langle T_\mathrm{air}\rangle-T_\mathrm{ice})
    \label{eq_Qice_field}
\end{align}
where $\alpha_\mathrm{ice}$ is the ice surface albedo, $\Phi(t)$ is the incoming solar radiation, $\langle T_\mathrm{air}\rangle$ is the mean air temperature, $T_\mathrm{ice}=273$~K is the melting ice temperature and $h_\mathrm{eff}$ is an empirical coefficient that has the dimension of an effective heat exchange coefficient. The data of $z_\mathrm{ice}(t)$ are shown in the suppl. mat. (see Fig.\figSupplConeFieldTime) and are used to determine the values of $\alpha_\mathrm{ice}$ and $h_\mathrm{eff}$ by adjusting the model of Eq.~\ref{eq_Qice_field}. A good overall agreement, given the simplicity of the model, is obtained with $\alpha_\mathrm{ice} = 0.32 \pm 0.02$ and $h_\mathrm{eff} = 14.8 \pm 0.5$~W$\cdot$K$^{-1}\cdot$m$^{-2}$, which is consistent with common values found in the literature for alpine glaciers~\cite{brock2000measurement, vincent_relative_2013}. As detailed in the suppl. mat., $h_\mathrm{eff}$ depends on the mean wind speed on the glacier which was constantly high in the period of interest.

\textbf{Early stage of the transition regime.} The heat flux received by the dirt layer can be split into two parts: one that depends on its surface temperature (less than what is received by bare ice if the dirt is warmer than $T_\mathrm{ice}$) and one received from the sun which depends on the dirt albedo $\alpha_\mathrm{dirt}$. %As we expect $\alpha_\mathrm{dirt}<\alpha_\mathrm{ice}$, the corresponding fraction of the heat flux can be amplified compared to bare ice. 
 The calculation leading to Eq.~\ref{eq_vt_vice_0} can be adapted for the daily averaged ratio of melt velocities in the presence of solar radiation:
\begin{align}
\left\langle \frac{v_\mathrm{t0}}{v_\mathrm{ice}} \right\rangle=\frac{1}{1+\mathrm{Bi}}\times \frac{1+(1-\alpha_\mathrm{dirt}) \langle \tilde \Phi \rangle}{1+(1-\alpha_\mathrm{ice}) \langle \tilde \Phi \rangle}
\label{eq_vt_vice_0_sun}
\end{align}
where $\langle \tilde \Phi \rangle = \langle \Phi(t) \rangle/(h_\mathrm{eff} (\langle T_\mathrm{air}\rangle-T_\mathrm{ice})) \approx 2.4$ is a dimensionless number accounting for the effect of solar radiation. This means that the albedo difference has only a correcting effect, and that the insulating effect acts on the total heat flux. This can be interpreted as follows: the solar incoming flux induces a strong thermal gradient across the dirt layer (whose bottom stays at $T_\mathrm{ice}$). This, in return, reduces (or even changes the sign, if the surface temperature is higher than $T_\mathrm{air}$) the other heat fluxes (wind induced, infrared, etc), ultimately reducing the heat flux received by the covered ice. 

The effective thermal conductivity of wet gravel collected on the Mer de glace was measured in the laboratory and found to be $\lambda_\mathrm{Gravel} = 0.73\pm 0.05$~W$\cdot$m$^{-1}\cdot$K$^{-1}$ (see supplementary materials). Given the mean effective heat exchange coefficient during the studied time period, this corresponds to a thermal length $\delta = 4.9 \pm 0.4$~cm. On the field data of Fig.~\ref{fig5}, an initial linear regime is clearly visible for $z_\mathrm{ice} < 0.2$~m, with a higher slope (corresponding to a more insulating behavior) for the thickest pile. Assuming that $e_\mathrm{t}\approx e_\mathrm{t0}$, Eq.~\ref{eq_vt_vice_0_sun} can be used to extract the last unknown parameter $\alpha_\mathrm{dirt}$ from these data. The best fitting value is $\alpha_\mathrm{dirt} = 0.20 \pm 0.05$ (see Fig.~\figSupplMesureDelta), which is compatible with values commonly used for gravel~\cite{PELTONIEMI2007434} or granite rock~\cite{WATSON197195, henot2022_TC}.

\textbf{Cone regime.} We assume that conditions (i-iii) of the previous model (subsection~\ref{subsection_cone_lab}) remain valid in the field. The hypothesis of volume conservation, however, needs to be adapted. Indeed, due to the cohesive nature of wet gravel and to the fact that the initial piles were compacted by hand, a decompaction can occur during the transition regime: the dirt volume $V$ in the cone regime is therefore larger than the initial volume $V_0$. We characterize this process by the parameter $f=V/V_0$. By comparing its dimensions in the initial and final states, we measured $f = 1.4 \pm 0.3$ for the cone shown in Fig.~\ref{fig1} (see suppl. mat.). We also noticed that the dirt covering natural cones on the glacier could easily be compacted by hand by $\approx 20-40$~\%. In the following we keep the assumption that the dirt volume is conserved throughout the cone regime, with the value $V=fV_0$. The melting velocity $v_\mathrm{s}$ on the side of the cone also has to be adapted in order to take into account the presence of solar radiation and the fact that the corresponding heat flux reaches the sides of the cone with an angle (averaged over a day and compared to a flat surface) which reduces the flux by a factor $\cos \theta$. This leads to:
\begin{align}
    \frac{\mathrm{d}h}{\mathrm{d}z_\mathrm{ice}} = 1-\frac{1}{\cos \theta}\frac{1}{1+Ae_\mathrm{t}(h)/\delta} \frac{1+(1-\alpha_\mathrm{dirt}) \langle \tilde \Phi \rangle \cos \theta}{1+(1-\alpha_\mathrm{ice}) \langle \tilde \Phi \rangle}
    \label{eq_dHdzice_sun}
\end{align}
By following the same steps as in subsection~\ref{subsection_cone_lab}, $H(z_\mathrm{ice})$ can be computed for each cone, as shown by solid lines in Fig.~\ref{fig5}. The computation uses the parameters $A$, $h_\mathrm{eff}$, $\alpha_\mathrm{ice}$, $\alpha_\mathrm{dirt}$, $\lambda_\mathrm{Gravel}$, $\langle \tilde \Phi \rangle$ and $\theta$ given previously, and $e_\mathrm{t0}$ and $V_0$ are given for each cone in Table~\ref{tableConesIni}. The only adjustable parameter here is the decompaction factor $f$, and the best fitting value was $f=1.4$. The main source of uncertainty on the model prediction, shown using a shaded area in Fig.~\ref{fig5}, is the inaccuracy on $V_0$ (15-20~\%). 

The beginning of the cone regime is well predicted for the thinnest cones (1 and 3) but is a bit premature for cone 2, which leads to a systematic underestimation of $H$. In the cone regime, the growth rate of the height is very well predicted for all three cases. Cone 2 starts with a dirt thickness about $3$ times higher than the fixed point of Eq.~\ref{eq_dHdzice_sun} : $e_\mathrm{t, \infty} \approx 3.4$~cm, which leads to a rapid growth in the cone regime as a lot of dirt will flow before the protective layer gets thin enough to reach the stationary regime. The model predicts a final height $H_\mathrm{\infty} \approx 57$~cm reached within 5~\% at $z_\mathrm{ice}\approx 2.2$~m. Cone 1 and 3 however start with a dirt thickness $e_\mathrm{t,0}$, close to $e_\mathrm{t, \infty}$, which explains why they do not grow much in the cone regime, as they have already almost reached their maximum height. 
%It appeared in this subsection that the simple model developed in a controlled environment could be adapted to quantitatively describe dirt cone formation on a glacier.

\textbf{Steady state.} As the cone height grows causing the dirt to flow, the cover gets thinner and less insulating. At some point $e_\mathrm{t}=e_\mathrm{t\infty}$, which corresponds to $v_\mathrm{t} = v_\mathrm{ice}$, and a steady state is reached. This final thickness is independent of the initial state and is fixed by the properties of the cover layer (thermal conductivity and mechanical properties) as well as the characteristics of the incoming heat (mainly $h_\mathrm{eff}$ and to a lesser extent $\langle \tilde \Phi \rangle$ and the ice and dirt albedo). It is worth noting that the final thickness, which controls the dynamics, can vary substantially over time on a glacier as $h_\mathrm{eff}$ depends on the average wind speed. The growth rate of a dirt cone can thus keep evolving even long after its formation. For example, a cone that forms and reaches a stationary state during a calm period (low $h_\mathrm{eff}$ and high $\delta$) will start growing again during a windy period (high $h_\mathrm{eff}$ and low $\delta$).
%\subsection{Physical mechanisms of cone formation}

%The good agreement between the various observations and the simple model described above allows to identify the physical mechanisms at play in the cone formation.  At some point $e_\mathrm{t}$ will reach $e_\mathrm{t~\infty}$ corresponding $v_\mathrm{t}$=$v_\mathrm{ice}$ and a stationary state will be reached. This final thickness is independent of the initial state and is only fixed by the properties of the dirt (thermal conductivity and mechanical properties) and by the effective heat exchange coefficient. This last parameter, which controls the dynamics, can vary substantially over time on a glacier as it depends on the wind mean speed. The growth rate of dirt cone can thus evolve even long after their formation. For example a cone that forms and reaches a stationary state during a calm period (low $h_\mathrm{eff}$ and high $\delta$) will start growing again during a windy period (high $h_\mathrm{eff}$ and low $\delta$).

\section{Conclusion and perspectives}\label{sec:ccl}
In this article, we described small-scale experiments reproducing dirt cone formation in a well controlled laboratory environment as well as time resolved observations of the formation of 3 cones on the Mer de glace. Dirt cone formation was also studied through 2D numerical simulations taking into account both the grain mechanics and the thermal heat exchanges that are able to reproduce well the cone formation process. A simple model was developed and led to a quantitative agreement with the laboratory and field observation as well as the simulations. This combination of approaches allowed us to gain insight into the physical mechanisms governing this structure formation. 

A dirt layer lying at the surface of a glacier acts as an insulation cover that reduces the ice melting under it. The differential ablation causes the ice surface to deform which induces a quasi-static flow of the dirt, starting from the edge of the pile. The structure acquires its conic shape when the deformation reaches the summit. The angle of the cone is determined by the mechanical properties of the grains (friction, cohesion) but does not correspond to a repose angle and is probably dependant on the history of stress distribution during the cone formation. As long as the dirt layer covering the cone is thick enough to reduce ice melting, the cone height will grow causing the dirt to creep along the sides, and get thinner and less insulating. Finally a stationary state is reached in which the insulating dirt cover exactly compensates for the fact that the structure received heat on a higher surface or with a lower albedo. In the model we developed, this final state is stable, which can explain the month-long lifetime of dirt cones on glaciers (while they typically form within a week).

%Our field observations, limited on the first 10 days of formation and to one time period, did not allow to test how the growth dynamics can be affected by the evolution of the meteorological conditions on the glacier and more especially the wind which is a prediction of our model. 
Our field observations along with the modeling open the possibility to use dirt cones as a proxy to estimate environmental parameters such as the heat exchange coefficient of the glacier ablation rate. On glaciers, the life-time of dirt cones is limited (to a few month) and the process by which this happens remains unexplained. It may be related to the progressive degradation of the dirt layer under the effect of rain or melt water, but clearly it deserves further attention. 
Another question that remains open is the formation of a "cone forest" observed on glaciers, consisting of several cones of various height, all in contact with each other. They clearly emerge from an initial large patch of dirt, but whether individual cones appear due to thickness inhomogeneity or from a more puzzling physical instability remains to be clarified. 

\medskip

\section*{Acknowledgements}

The authors are grateful to Marine Vicet, Jérémy Vessaire and Thierry Dauxois for fruitful discussions, to the Fédération de Recherche Marie André Ampère and to the Laboratoire de Physique at the ENS de Lyon for ﬁnancial support.

\section*{Appendix: volume conservation}

In the model developed in the main text, one important assumption is the volume conservation of the granular cover. In the following we detail the corresponding calculation.

In the cone regime, the dirt cone is assumed to have a constant thickness $e_\mathrm{s} = A e_\mathrm{t}$. In 2D, the dirt layer has a length $\approx (h+e_\mathrm{t}/2)/\sin \theta$ on each side. The volume per unit depth can thus be approximated as:
\begin{align}
\mathcal{V}_\mathrm{2D} = \frac{A}{\sin \theta}(2h + e_\mathrm{t})e_\mathrm{t}
\end{align}
This quantity is assumed to be constant $\mathrm{d}\mathcal{V}_\mathrm{2D}=0$ leading to the following differential equation:
\begin{align}
1 + \frac{h}{e_\mathrm{t}} + \frac{\mathrm{d}h}{\mathrm{d}e_\mathrm{t}} = 0
\end{align}
whose solution is:
\begin{align}
h(e_\mathrm{t}) = \frac{e_\mathrm{t0}}{e_\mathrm{t}}\left(h_1+\frac{e_\mathrm{t0}}{2}\right) - \frac{e_\mathrm{t}}{2}
\end{align}
which can be inverted in:
\begin{align}
e_\mathrm{t~2D}(h) = \sqrt{h^2+2e_\mathrm{t0}\left(h_1+\frac{e_\mathrm{t0}}{2}\right)} - h
\end{align}
In 3D, the same approximation leads to a volume:
\begin{align}
\mathcal{V}_\mathrm{3D} = \frac{\pi A}{\sin\theta\tan\theta}e_\mathrm{t}\left(h+\frac{e_\mathrm{t}}{2}\right)^2
\label{eq_suppl_v03d}
\end{align}
The condition $\mathrm{d}\mathcal{V}_\mathrm{3D}=0$ corresponds to:
\begin{align}
\frac{3}{2} + \frac{h}{e_\mathrm{t}} + 2\frac{\mathrm{d}h}{\mathrm{d}e_\mathrm{t}} = 0
\end{align}
whose solution is:
\begin{align}
h(e_\mathrm{t}) = \frac{h_1\sqrt{e_\mathrm{t0}} +e_\mathrm{t0}^{3/2}/2}{\sqrt{e_\mathrm{t}}} - \frac{e_\mathrm{t}}{2}
\end{align}
The expression of $e_\mathrm{t~3D}(h)$ is then obtained by keeping the only real solution of the cubic equation with unknown $x=\sqrt{e_\mathrm{t}}$.

The volume of the initial pile in 2 and 3D is:

\begin{align}
\mathcal{V}_\mathrm{2D~0} = \left(R_0-\frac{e_\mathrm{t0}}{\tan \theta_0}\right)e_\mathrm{t0}
\\
%\quad, \qquad
\mathcal{V}_\mathrm{3D~0} = \frac{\pi}{3}\tan \theta_0\left[R_0^3-\left(R_0-\frac{e_\mathrm{t0}}{\tan \theta_0}\right)^3\right]
\label{eq_suppl_vcone}
\end{align}
If the volume is assumed to be conserved from the initial flat pile state to the cone regime, the parameter $h_1=h(e_\mathrm{t0})$ can be expressed as:
\begin{align}
h_{2D~1} = \left(R_0-\frac{e_\mathrm{t0}}{2\tan \theta_0}\right)\frac{\sin \theta}{A} - \frac{e_\mathrm{t0}}{2}
\label{eq_Ap_h1_2D}
\\
%\quad, \qquad
h_\mathrm{3D~1} = \sqrt{\frac{\mathcal{V}_0\sin\theta\tan\theta}{A\pi e_\mathrm{t0}}} - \frac{e_\mathrm{t0}}{2}
\label{eq_Ap_h1_3D}
\end{align}

\bibliographystyle{apsrev4-2}
\bibliography{bibliographie}

\clearpage
\onecolumngrid
\renewcommand\thefigure{S-\arabic{figure}}    
\renewcommand\thetable{S-\Roman{table}}  
\renewcommand{\theequation}{S-\arabic{equation}}
\setcounter{equation}{0}
\setcounter{figure}{0}
\setcounter{table}{0}
\setcounter{section}{0}

\begin{center}
    \large \textbf{Supplementary materials}
\end{center}

\section{Thermal length measurement in the lab}

\begin{figure}[htbp]
  \centering
  \includegraphics[width=10cm]{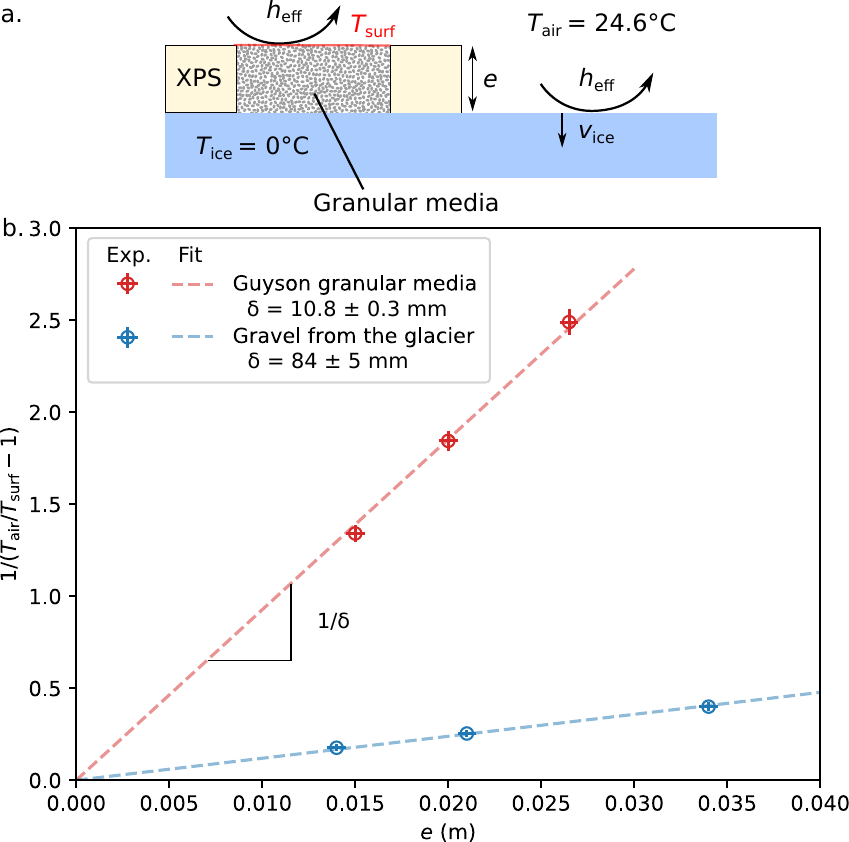}
 \caption{(a) Principle of determination of the thermal length $\delta$ of a granular material in contact with ice. (b) $1/(T_\mathrm{air}/T_\mathrm{surf}-1)$ as a function of the thickness $e$ of granular materials for plastic grains (red markers) and gravel (blue markers). The dashed lines are linear adjustments with slope $1/\delta$.}
  \label{suppl_mesure_delta}
\end{figure}

The thermal length $\delta=\lambda/h_\mathrm{eff}$ in the lab was measured for the Guyson plastic grains and for the gravel constituting the natural dirt cones on the Mer de glace glacier. The measurement principle is illustrated in Fig.~\ref{suppl_mesure_delta}a. Cylindrical containers of radius 2.5~mm, and height in the range $e = 13-34$~mm, made of extruded polystyrene (XPS) were filled with the granular medium, initially dry. The bottom surface consists of an ice block at temperature $T_\mathrm{ice} = 0$~\degree C and the top surface is open to the room atmosphere at temperature $T_\mathrm{air} = 24.6$~\degree C. The gravel is quickly wetted by the capillary ascension of melt water while the plastic grains, made hydrophobic stay dry. The system is let to thermally equilibrate for 30~min. The surface temperature $T_\mathrm{surf}$ is then measured using an IR camera. This measurement is done for different thicknesses $e$. As the thermal conductivity of XPS is two orders of magnitude lower than the one of the granular material, the problem can be assumed 1D with a vertical flux $j_\mathrm{ice \rightarrow air} = \lambda (T_\mathrm{surf}-T_\mathrm{ice})/e = h_\mathrm{eff} (T_\mathrm{air}-T_\mathrm{surf})$. From this, the differences in temperature can be related to the thickness $e$ by :
\begin{align}
\frac{T_\mathrm{surf}-T_\mathrm{ice}}{T_\mathrm{air}-T_\mathrm{surf}} = \frac{1}{\delta} e
\end{align}
Fig.~\ref{suppl_mesure_delta}b shows $1/(T_\mathrm{air}/T_\mathrm{surf}-1)$ as a function of $e$ which are expected to be proportional with slope $1/\delta$. A linear regression leads to $\delta = 10.8 \pm 0.3$~mm for the plastic grains and $\delta=84\pm 5$~mm for the gravel. The effective heat exchange coefficient, assumed to be identical for the granular material and the ice surface, was measured by following the melt velocity of the ice block $v_\mathrm{ice} = (7.1 \pm 0.2) \times 10^{-7}$~m$\cdot$s$^{-1}$ in the same environment using a camera watching from the side. This leads to $h_\mathrm{eff} = v_\mathrm{ice}\mathcal{L}_\mathrm{ice}/(T_\mathrm{air}-T_\mathrm{ice}) = 8.7\pm 0.3 $~W$\cdot$K$^{-1}\cdot$m$^{-2}$ and thus to $\lambda_\mathrm{wet~plastic~grains} = 0.094 \pm 0.005 $~W$\cdot$m$^{-1}\cdot$K$^{-1}$ and $\lambda_\mathrm{wet~gravel} = 0.73 \pm 0.05 $~W$\cdot$m$^{-1}\cdot$K$^{-1}$.

%%%%%%%%%%%%%%%%%%%%%%%%%%%%%%%%%%%%%%%%
%%%%%%%%%%%%%%%%%%%%%%%%%%%%%%%%%%%%%%%%

%\section{Details on the model: conservation of volume of the granular media} 

\section{2D numerical simulations - Effect of the friction coefficient}

In order to avoid crystallisation, a $\pm 20 \%$ polydispersity was introduced on the radii distribution of the grain around a mean value $\langle r \rangle$.

%%%%%%%%%%%%%%%%%%%%%%%%%%%%%%%%%%%%%%%%
\begin{table*}%  table1,  one  column
\label{table_simu_params}
\begin{ruledtabular}
\begin{tabular}{cccc}
 Mean grain radius & $\langle r \rangle$  & $2.5\times 10^{-4}$ & m \\
 Grain radius polydispersity & & $20$ & $\%$ \\
 Grain specific density &  & 2 & kg$\cdot$m$^{-2}$\\
 Time step & $dt$ & $5\times 10^{-7}$ & s \\
 Ice melting velocity & $v_\mathrm{ice}$ & $1\times 10^{-3}$ & m$\cdot$s$^{-1}$ \\
 Gravitational acceleration &$g$ & 10 & m$\cdot$s$^{-2}$\\
 Normal spring constant & $k_\mathrm{n}$ & $1\times10^{3}$ & N$\cdot$m$^{-1}$\\
 Tangential spring constant & $k_\mathrm{t}$ & $0.285 k_\mathrm{n}$ & N$\cdot$m$^{-1}$ \\
 Dashpot constant & $\gamma$ & 0.01 & N$\cdot$s$\cdot$m$^{-1}$ \\
 Friction coefficient & $\mu$ & 0.3 and 0.6 & \\
 Cohesion force (dimers) & & $7\times10^{-2}$ & N \\
 Rotational spring constant (dimers) & & $6.25\times10^{-4}$ & N$\cdot$m$\cdot$rad$^{-1}$ \\
 \hline
  Effective thermal conductivity of the granular media & $\lambda_\mathrm{eff}$ & 0.18 & W$\cdot$m$\cdot^{-1}$K$^{-1}$ \\
 Effective heat exchange coefficient & $h_\mathrm{eff}$ & 9 & W$\cdot$m$\cdot^{-2}$K$^{-1}$ \\
 Air temperature & $T_\mathrm{air}$ & 21.5 & \degree C \\
 Ice temperature & $T_\mathrm{ice}$ & 0 & \degree C \\
\end{tabular}
\end{ruledtabular}
\caption{Parameters of the 2D numerical simulation.}
\end{table*}
%%%%%%%%%%%%%%%%%%%%%%%%%%%%%%%%%%%%%%%%

%%%%%%%%%%%%%%%%%%%%%%%%%%%%%%%%%%%%%%%%
\begin{figure}[htbp]
  \centering
  \includegraphics[width=\linewidth]{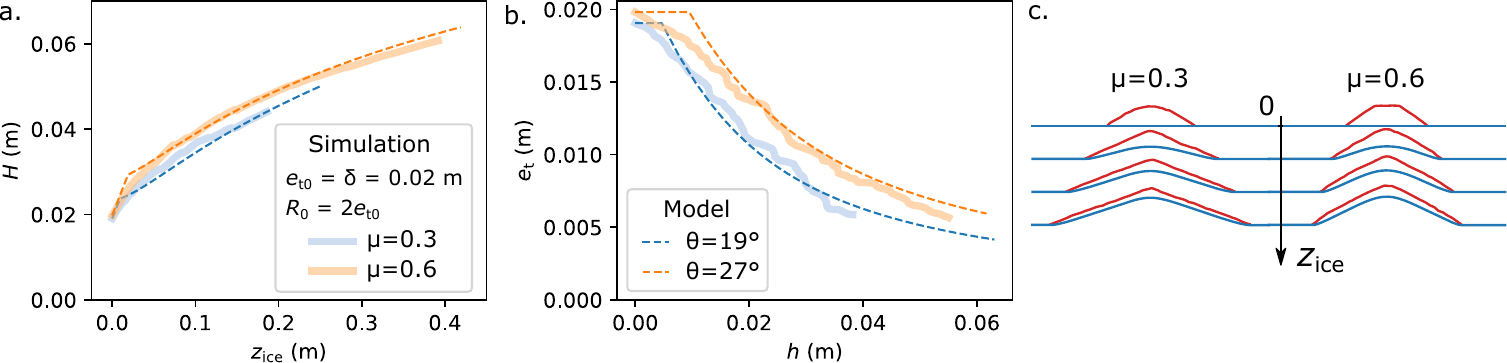}
 \caption{Results of 2D numerical simulations (solid lines), for $\mu=0.3$ and $\mu=0.6$, showing the total cone height $H$ as a function of the ablated ice thickness $z_\mathrm{ice}$ (a) and the granular thickness as the top of the cone $e_\mathrm{t}$ as a function of its height $h$ (b). The model is plotted in dashed line with $\theta_0$ depending on $\mu$ and with adjustable parameter $\theta$. (c). Profile evolution of the ice (blue) and of the top of the grain layer (red).}
  \label{suppl_simu_modele_mu}
\end{figure}
%%%%%%%%%%%%%%%%%%%%%%%%%%%%%%%%%%%%%%%%

The effect of the inter grain friction coefficient $\mu$ in the 2D numerical simulation was studied for one initial configuration ($e_\mathrm{t0}=\delta$, $R_0 = 2\delta$). The results are shown in Fig.~\ref{suppl_simu_modele_mu} for $\mu=0.3$ and $\mu=0.6$. As expected, the angle of the initial pile $\theta_0$, which corresponds to a repose angle depends on $\mu$: $\theta_0(\mu=0.3) = 31\pm 2$\degree~ and $\theta_0(\mu=0.6) = 36\pm 2$\degree~. More interestingly it is also the case of the cone angle measured on the profiles (see Fig.~\ref{suppl_simu_modele_mu}c): $\theta(\mu=0.3) = 19\pm 2$\degree and $\theta(\mu=0.6) = 26\pm 2$\degree. The model described in the main text was adjusted on these results with $\theta$ kept as adjustable parameter (due to the high sensibility of the model on this parameter) leading to a good agreement for $\theta$ values compatible with the measured one.

\section{Details on the field observations}

%%%%%%%%%%%%%%%%%%%%%%%%%%%%%%%%%%%%%%%%
\begin{figure*}[htbp]
  \centering
  \includegraphics[width=\linewidth]{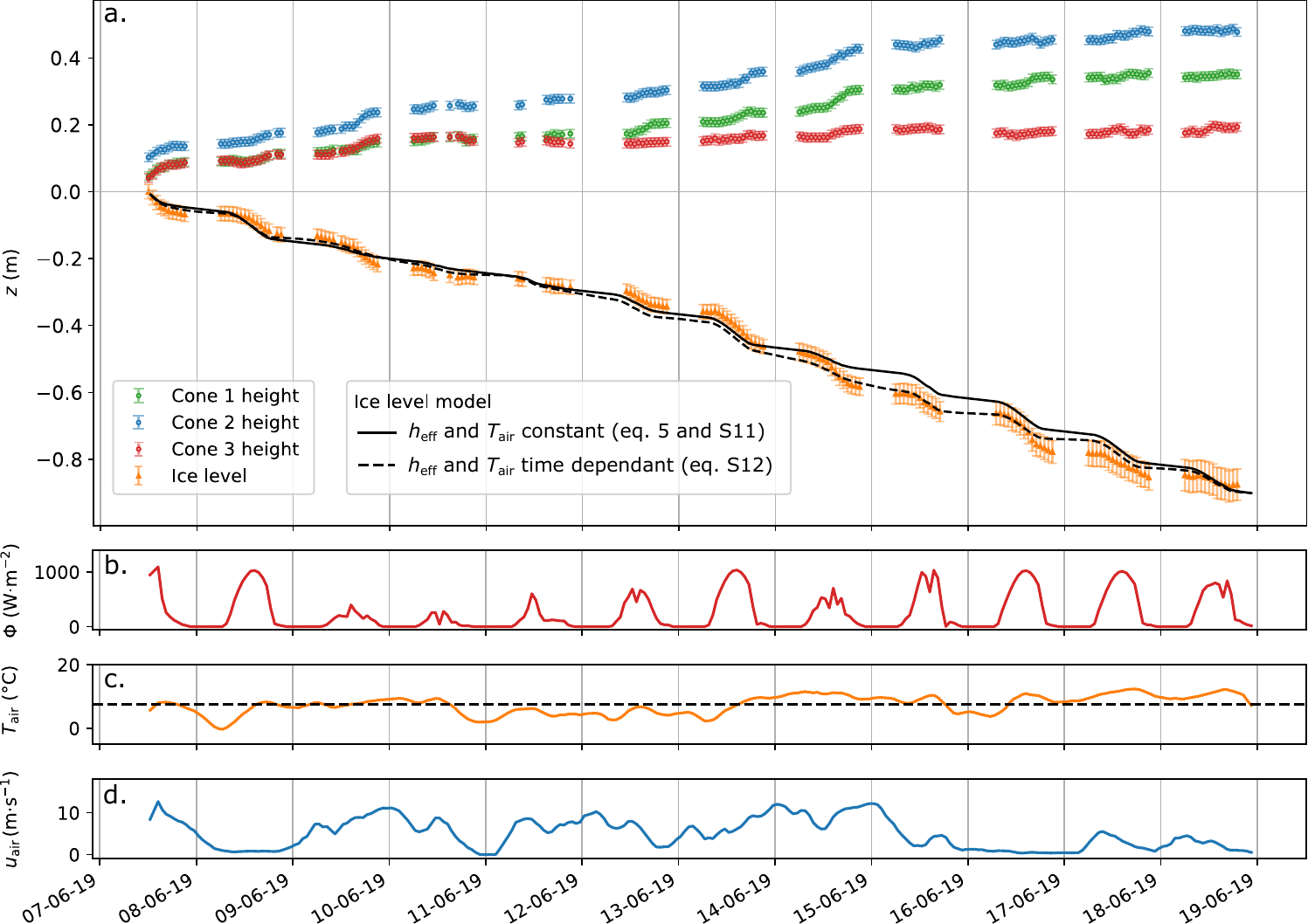}
 \caption{(a) Evolution of the heights of the tree cones (circles) and of the level of the ice (triangle) between the 7 and the 19 June 2019. The black solid line corresponds to the ice ablation computed from the simple model used in the main text (Eq.~\ref{eq_Qice_field_suppl} with $\alpha_\mathrm{ice}=0.32$ and $h_\mathrm{eff} = 14.5$~W$\cdot$K$^{-1}\cdot$m$^{-2}$) while the dashed line corresponds to a more sophisticated model (Eq.~\ref{eq_Qice_field_suppl2} with $\alpha_\mathrm{ice}=0.32$, $h_\mathrm{0} = 3.2$~W$\cdot$K$^{-1}\cdot$m$^{-2}$ and $\beta=2.5$~J$\cdot$K$^{-1}\cdot$m$^{-3}$).
 (b-d) Meteorological data measured at the Requin AWS: solar radiative flux $\Phi$ (b),  air temperature $T_\mathrm{air}$ (c) (The dashed line corresponds to $\langle T_\mathrm{air}\rangle$), wind speed $u_\mathrm{a}$ (d).}
  \label{suppl_cone_field_time}
\end{figure*}
%%%%%%%%%%%%%%%%%%%%%%%%%%%%%%%%%%%%%%%%

\subsection{Evaluation of $z_\mathrm{ice}$ and $H$}
The field data for the 3 cone were manually extracted from the time lapse pictures. As shown in the video ? in the suppl. mat., the top and the right and left bottom of each structure were pointed for each frame. The altitude $z_\mathrm{ice}$ of the ice surface was deduced from the vertical position of the 2 bottom points of each structures. The 3 dataset coincided ($\pm 1$~cm) during the first 6 days (7-12/06/2019) and then progressively shifted due to the distortion of the images induced by the movement of the camera ($\pm 4$~cm on the 15/06/2019, $\pm 13$~cm on the 18/06/2019). In order to minimize this effect, $z_\mathrm{ice}$ was obtained by averaging over the 3 dataset (see orange markers in Fig.~\ref{suppl_cone_field_time}a). The height $H$ of each structure was evaluated from the difference in the vertical axis between the top point and the mean of the two bottom points (see green, blue and red markers in Fig.~\ref{suppl_cone_field_time}a).

\subsection{Ice melting models, determination of $\alpha_\mathrm{ice}$ and $h_\mathrm{eff}$}
The solar heat flux $\Phi(t)$ and air temperature $T_\mathrm{air}(t)$ were measured at the Requin AWS situated 600~m higher and 3 km away from the measurement site (see~\cite{henot2022_TC} and its supplements for location on map and details on the temperature correction) and are shown in Fig.~\ref{suppl_cone_field_time}b and c. 

The ice melting model described in the main text (Eq.~5) is an enhanced temperature model in which the air temperature is replaced by its mean value over the time period:
\begin{align}
    Q_\mathrm{env.\rightarrow ice}= (1-\alpha_\mathrm{ice})\Phi(t) + h_\mathrm{eff}[\langle T_\mathrm{air}\rangle-T_\mathrm{ice}]
    \label{eq_Qice_field_suppl}
\end{align}
In the glaciology literature, the heat exchange coefficient $h_\mathrm{eff}$ is usually called a temperature factor $TF$ and expressed as a height of melt water (water equivalent: w.e.) per unit time and per unit temperature: $TF = h_\mathrm{eff} \rho_\mathrm{water} / (\mathcal{L}_\mathrm{fus} \rho_\mathrm{ice})$ where $\rho_\mathrm{water}$ and $\rho_\mathrm{ice}$ are the specific density of water and ice respectively.
The integration of this equation was adjusted to the observation of $z_\mathrm{ice}(t)$ as shown using a solid black line in Fig.~\ref{suppl_cone_field_time}a with best fitting value of parameters $\alpha_\mathrm{ice} = 0.32$ and $h_\mathrm{eff} = 14.8$~W$\cdot$K$^{-1}\cdot$m$^{-2}$ ($TF = 4.7 \pm 0.2 $~mm w.e.$\cdot$day$^{-1}\cdot$K$^{-1}$). The overall agreement with the data is acceptable given the simplicity of the model. These parameters values are also compatible with the literature: $\alpha_\mathrm{ice} = 0.10-0.35$ for ice weathered surface in the ablation zone of Haut Glacier d'Arolla, Switzerland~\cite{brock2000measurement} and $TF = 4.5-6.2 $~mm w.e.$\cdot$day$^{-1}\cdot$K$^{-1}$ with a daily averaged enhanced temperature index model applied over 7 years at Glacier de Saint-Sorlin, France~\cite{ vincent_relative_2013}.

In order to emphasize the fact that this model captures the main physical source of ice melting, a slightly more refined model was tested with time dependant air temperature $T_\mathrm{air}(t)$ and effective exchange coefficient $h_\mathrm{eff}(t)$ (depended on the wind speed $u_\mathrm{air}(t)$):
\begin{align}
    Q_\mathrm{env.\rightarrow ice}= (1-\alpha_\mathrm{ice})\Phi(t) + h_\mathrm{eff}(u_\mathrm{air}(t))[ T_\mathrm{air}(t)-T_\mathrm{ice}], \qquad 
    h_\mathrm{eff}(u_\mathrm{air}(t)) = h_0 + \beta u_\mathrm{air}(t)
    \label{eq_Qice_field_suppl2}
\end{align}
This later point is justified by the fact that most of the non-solar incident heat flux comes from turbulent flux (sensible and latent) which is proportional to the wind speed $u_\mathrm{air}(t)$ which was measured at the Requin AWS and plotted in Fig.~\ref{suppl_cone_field_time}d. The integration of Eq.~\ref{eq_Qice_field_suppl2} is shown in Fig.~\ref{suppl_cone_field_time}a in dashed black line with best fitting parameters $\alpha_\mathrm{ice} = 0.32$, $h_\mathrm{0} = 3.2$~W$\cdot$K$^{-1}\cdot$m$^{-2}$ and $\beta=2.5$~J$\cdot$K$^{-1}\cdot$m$^{-3}$ corresponding to the same mean value than the previous model: $\langle h_0 + \beta u_\mathrm{air}(t) \rangle = 14.8$~W$\cdot$K$^{-1}\cdot$m$^{-2}$ and leading to a slightly better agreement. For the sake of simplicity we choose to keep constant values of $T_\mathrm{air}$ and $h_\mathrm{eff}$ in the cone formation model developed in the main text although this refinement could be taken into account in the model without much change. 

\subsection{Beginning of the transition regime, determination of $\alpha_\mathrm{gravel}$}

\begin{figure*}[htbp]
  \centering
  \includegraphics[width=13cm]{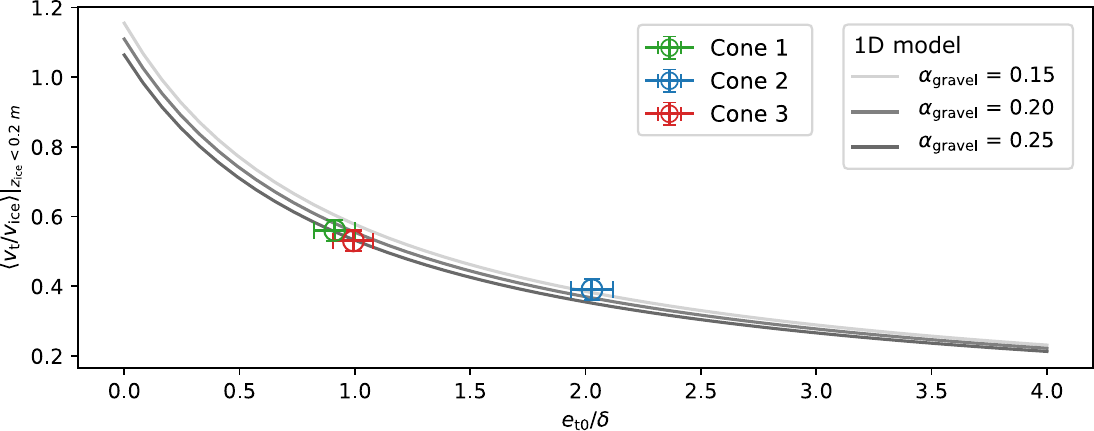}
 \caption{Mean ratio $\langle v_\mathrm{t}/v_\mathrm{ice}\rangle$ corresponding to the slope of $H(z_\mathrm{ice})$ for $z_\mathrm{ice}<0.2$~m (see Fig.~\ref{fig5}) determined from a linear regression for the 3 cones (markers), as a function of the $e_\mathrm{t0}/\delta$ where $\delta=\lambda_\mathrm{gravel}/\langle h_\mathrm{eff}\rangle$ with $\lambda_\mathrm{gravel} = 0.73 \pm 0.05 $~W$\cdot$m$^{-1}\cdot$K$^{-1}$ and $\langle h_\mathrm{eff}\rangle = 14.8 \pm 0.5$~W$\cdot$K$^{-1}\cdot$m$^{-2}$. The solid lines corresponds to a 1D conduction model (Eq.~\ref{eq_vt_vice_0_sun} of the main text) for three values of $\alpha_\mathrm{dirt}=\alpha_\mathrm{gravel}$.}
  \label{suppl_field_initial}
\end{figure*}

At the beginning of the transition regime, the dirt piles are still flat around their center and only the sides are deformed. The dirt thickness $e_\mathrm{t} \approx e_\mathrm{t0}$ can be asssumed constant and the thermal problem can be considered 1D. This lead to Eq.~\ref{eq_vt_vice_0_sun} of the main text and predict a regime of constant slope for $H(z_\mathrm{ice}) = (1-\frac{v_\mathrm{t}}{v_\mathrm{ice}})z_\mathrm{ice}$. This is indeed what we see in Fig.~\ref{fig5} for $z_\mathrm{ice}<0.2$~m. From a linear adjustment of the data, we deduce the slope $\langle\frac{v_\mathrm{t}}{v_\mathrm{ice}}\rangle$ for each structure. These quantities are plotted in Fig.~\ref{suppl_field_initial} as a function of $e_\mathrm{t0}/\delta$ with $\delta = 4.9 \pm 0.4$~cm. Eq.~\ref{eq_vt_vice_0_sun} is also plotted in solid line for $\langle \tilde{\Phi}\rangle$=2.4, $\alpha_\mathrm{ice}=0.32$ and 3 values of $\alpha_\mathrm{dirt}$ from 0.15 to 0.25 which is the last unknown parameter. These values leads to a good agreement with the experimental point given the uncertainty. We thus estimate this parameter as $\alpha_\mathrm{dirt} = 0.20 \pm 0.05$. The dirt here is made of granite gravel of millimetric size. This value is compatible with the literature: the reflectance of grey granite gravel was measured in~\cite{PELTONIEMI2007434} who reported values ranging from 0.14 to 0.20 at 700~nm. The reflectance of granite rocks was measured in~\cite{WATSON197195} which after integration over the solar spectrum gives an albedo of 0.18 (see suppl. mat. of \cite{henot2022_TC}).

%%%%%%%%%%%%%%%%%%%%%%%%%%%%%%%%%%%%%%%%
%%%%%%%%%%%%%%%%%%%%%%%%%%%%%%%%%%%%%%%%

\subsection{Volume conservation of the cone studied in 2021}
The cone shown in Fig.~\ref{fig1} and~\ref{fig2} was made on the Mer de glace on 03/06/2021. The initial pile had a radius $R_0 = 30 \pm 1$~cm, a thickness $e_\mathrm{t0} = 9.5\pm 0.5$~cm and an angle $\theta_0 = 55 \pm 5$~\degree corresponding to an initial volume of dirt computed from Eq.~\ref{eq_suppl_v03d} of $V_0 = 21 \pm 3$~L. On 17/06/2021, pictures of this cone were taken before and after removing the dirt cover (see Fig.~\ref{fig1} and \ref{suppl_A}c) allowing to measure $H = 56 \pm 2$~cm, $e_\mathrm{t} = 3.5 \pm 0.2$~cm and the cone angle $\theta = 46 \pm 3$\degree (slightly less steep than the cone studied in 2019). The value of $A=0.56\pm0.02$ was also measured for this specific cone (see section~\ref{suppl_section_A} of the suppl. mat.). The final dirt volume can thus be computed from these data using Eq.~\ref{eq_suppl_vcone}: $V = 30 \pm 5$~L. The decompaction factor can thus be estimated to $f = 1.4 \pm 0.3$.

\section{Determination of the coefficient $A$} 
\label{suppl_section_A}

%%%%%%%%%%%%%%%%%%%%%%%%%%%%%%%%%%%%%%%%
\begin{figure}[htbp]
  \centering
  \includegraphics[width=12cm]{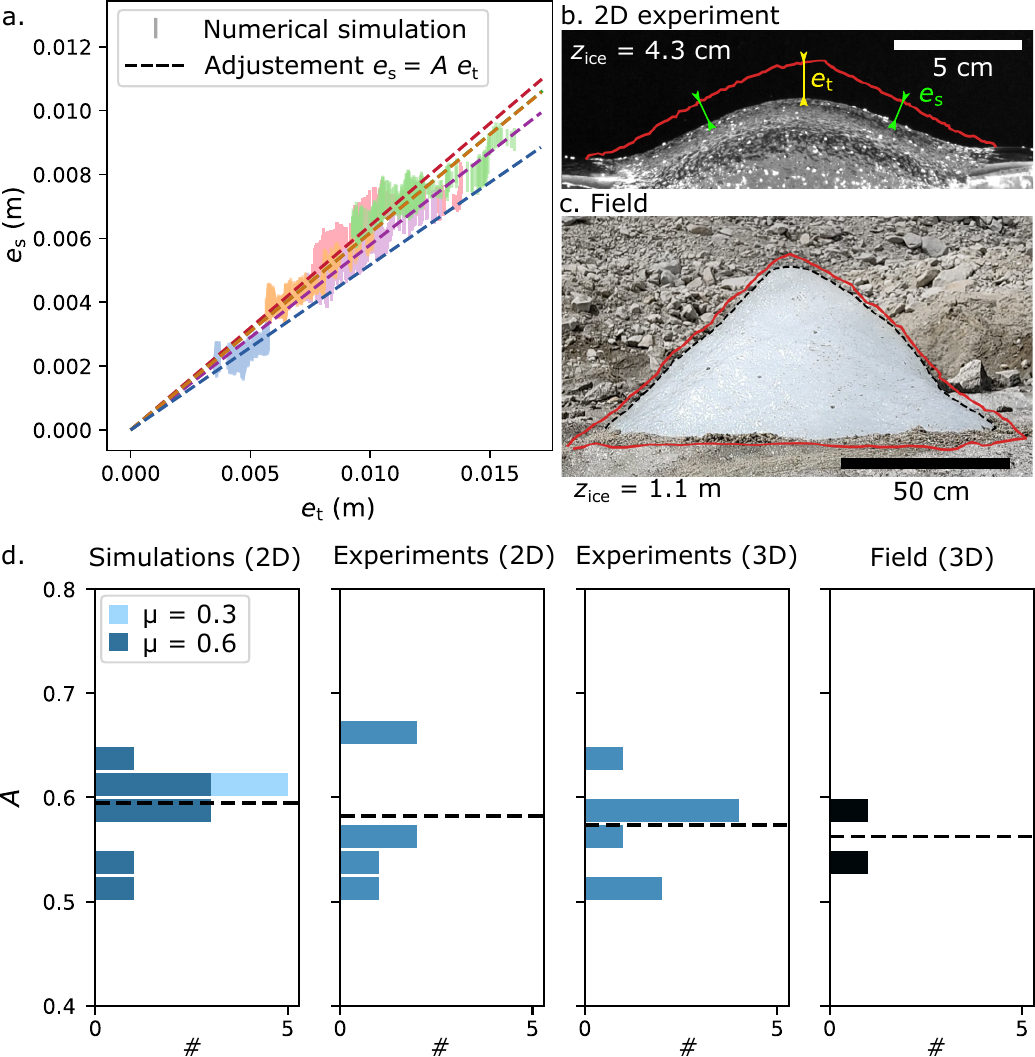}
 \caption{(a) Granular layer thickness $e_\mathrm{s}$ on the side of the cones as a function as the thickness at the top $e_\mathrm{t}$ for 2D simulations in the cone regime ($z_\mathrm{ice}>0.1$~m). The dashed lines are linear adjustments with slope $A$. (b,c) Determination of $e_\mathrm{t}$ and $e_\mathrm{s}$ after removal of the granular for a lab-controlled experiment (b) and on the field (c). (d) Distribution of the $A$ values for 2D simulations, 2D and 3D lab-controlled experiments and from a measurement performed on the field. The dashed lines correspond to the mean values.}
  \label{suppl_A}
\end{figure}
%%%%%%%%%%%%%%%%%%%%%%%%%%%%%%%%%%%%%%%%

An important assumption in the cone formation model developed in the main text is the fact that the ratio $A=e_\mathrm{s}/e_\mathrm{t}$ is a constant. This is far from obvious \textit{a priori} and results from observations. Fig.~\ref{suppl_A}a represents  $e_\mathrm{s}$ as a function of $e_\mathrm{t}$ in the cone regime for each numerical simulation. We see that for each run, these two quantity appear proportional to each other. The dashed lines are linear fit of these data with slope $A$. The distribution of these values are plotted on the histogram of Fig.~\ref{suppl_A}d. This ratio was also measured for all final state of the lab experiment (on the 2 sides) although only the beginning of the cone regime was reached. Fig.~\ref{suppl_A}b illustrates how this measurement was performed and the distribution of $A$ is shown on the histograms of Fig.~\ref{suppl_A}d for 2D and 3D experiments. Finally, this ratio was measured on the field for the cone shown in Fig.~\ref{fig1} of the main text using the two pictures taken before and after cleaning the dirt on the cone allowing to extract dirt and ice profiles as shown in Fig.~\ref{suppl_A}c in red and black. From these the top dirt thickness $e_\mathrm{t} = 3.50 \pm 0.10$~cm and the side dirt thickness on both sides $e_\mathrm{s,left} = 2.05 \pm 0.10$~cm, $e_\mathrm{s,right} =1.90 \pm 0.10$~cm could be measured. The corresponding values of $A$ are shown on the histogram of Fig.~\ref{suppl_A}d. It is striking that each of these cones in various configuration (2/3D, angle varying from 19 to 49\degree) display such a narrow distribution of the ratio $A$ with a mean $\langle A \rangle = 0.6$ and all values lying between $0.5$ and $0.7$.

For $A$ to stay constant, both the ice and the grain profiles have to evolve in a coupled way when the dirt thickness diminishes. Indeed, the top dirt thickness $e_\mathrm{t}$ is fixed by the shape taken by the ice profile when the cone forms and by the shape of the top of the grain layer that has to do with its stability. The coupling results from the complex interaction between the quasi static granular flow which controls the gain top profile and the thermal flux across the dirt layer which controls the gain bottom profile.

\end{document}